\documentclass[11pt]{article}

\usepackage{amssymb}
\usepackage{esvect}
\usepackage{latexsym}
\usepackage[tbtags]{amsmath}
\usepackage{amscd}
\usepackage{multirow}
\usepackage{amsthm}
\usepackage{color}
\usepackage{enumerate}
\usepackage{bm}
\usepackage{bbm}
\usepackage{setspace}
\usepackage{adjustbox} 
\usepackage{subfigure} 
\usepackage{lipsum}
\usepackage{mathtools}
\usepackage[titletoc,title]{appendix}
\usepackage[T1]{fontenc}
\usepackage{dsfont}
\usepackage{natbib}
\usepackage{url}
\usepackage[font=small,labelfont=bf]{caption}
\usepackage{authblk} 
\usepackage[hyperfootnotes=false]{hyperref} 
\hypersetup{
	colorlinks = true,
	linkcolor = blue, 
	filecolor = magenta,
	urlcolor = blue, 
	citecolor = blue 
}
\usepackage{cleveref}
\usepackage{sectsty}
\usepackage{fullpage}
\usepackage[symbol,multiple]{footmisc}
\usepackage{kotex}
\usepackage{graphicx}
\graphicspath{{figs/}}

\newcommand{\indep}{\perp \!\!\! \perp}

\newtheorem{definition}{Definition}

\newtheorem{proposition}{Proposition}

\allowdisplaybreaks
\usepackage{tcolorbox}
\usepackage{booktabs}

\newcommand{\beginsupplement}{%
	\setcounter{section}{0}%
	\renewcommand{\thesection}{S\arabic{section}}%
	\setcounter{figure}{0}%
	\renewcommand{\thefigure}{S\arabic{figure}}%
	\setcounter{table}{0}%
	\renewcommand{\thetable}{S\arabic{table}}%
}

\title{Weighted Conformal Prediction for Survival Analysis\\under Covariate Shift}

\author[1]{Jaeyoung Shin}
\author[1,2]{Chi Hyun Lee\footnote{email: chihyunlee@yonsei.ac.kr}}
\author[1,2]{Sangwook Kang}
\affil[1]{Department of Statistics and Data Science, Yonsei University, Seoul, South Korea}
\affil[2]{Department of Applied Statistics, Yonsei University, Seoul, South Korea}

\date{}

\begin{document}
	
\maketitle

\begin{abstract}
\noindent Reliable uncertainty quantification is essential in survival prediction, particularly in clinical settings where erroneous decisions carry high risk. Conformal prediction has attracted substantial attention as it offers a model-agnostic framework with finite-sample coverage guarantees. Extending it to right-censored outcomes poses nontrivial challenges. Several adaptations of conformal approaches for survival outcomes have been developed, but they either rely on restrictive censoring settings or substantial computation. A recent conformal approach for right-censored data constructs censoring-adjusted p-values and enables prediction intervals in general survival settings. However, the empirical coverage depends sensitively on heuristic tuning choices and its validity is limited to scenarios without covariate shift. In this paper, we establish theoretical justification for its prediction-set construction, providing a principled basis for defining prediction-set bounds, and extend the approach to covariate-shift settings. Simulation studies and a real data application demonstrate that the proposed method achieves robust coverage and coherent interval structure across varying censoring levels and covariate-shift settings.

\vspace{0.5cm}
Keywords: Conformal inference, nonconformity score, prediction interval, quasi-concavity, random censoring
\end{abstract}

\section{Introduction}\label{sec1}

In modern statistical learning, predictive accuracy and flexibility in modeling complex relationships between predictors and outcome have been the central objectives. However, in high-stakes applications such as clinical decision making, where the consequences of erroneous predictions can be severe, reliable uncertainty quantification is essential. Accordingly, prediction intervals have been extensively studied to measure the credibility of predictions \citep{geisser1993, krishnamoorthymathew2009}. Recently, conformal prediction has gained considerable attention as a model-agnostic framework that provides finite-sample coverage guarantees \citep{vovketal2005, shafer2008, leietal2018}.

While conformal prediction is appealing, extending the framework to survival outcomes is nontrivial due to right-censoring. Because censoring hides the actual event time for some individuals, censored and uncensored observations carry different amounts of information. This breaks the exchangeablity condition required for standard conformal prediction, which has motivated several adaptations of conformal prediction for censored data.

\cite{candesetal2023} and \cite{guietal2024} developed a conformal inference based method that produce lower prediction bounds under the Type I censoring setting by discarding observations with early censoring and applying reweighting to correct for subsetting. However, censoring times are not always observed for all individuals in general and truncating the outcome complicates interpretation when prediction of the actual event time is needed. Under general right-censoring, \cite{qinetal2025} proposed a resampling approach that calibrates conformity scores by approximating the sampling distribution of the conditional survival probability, which enables both one- and two-sided prediction intervals but requires correct model specification and substantial computation. \cite{yietal2025} introduced survival conformal prediction (SCP) for general right censoring by developing a censoring-adjusted conformity score and corresponding p-values. This provides a valuable step toward valid, model-agnostic inference under censoring, yet several theoretical and practical aspects remain insufficiently understood. In particular, its empirical coverage depends sensitively on heuristic tuning choices (e.g., censoring thresholds and candidate time grids).


In this paper, we address two key limitations of SCP: the lack of a theoretical understanding of its p-value structure and the need to extend its validity to settings where the training and test covariate distributions may differ. We first establish the quasi-concavity of SCP p-values, providing a principled basis for constructing prediction-set bounds. We then introduce a weighted extension of SCP that maintains this structural property and achieves valid coverage under covariate shift. The remainder of the paper is structured as follows. Section~\ref{sec:scp} reviews the SCP framework and establishes its theoretical properties. Section~\ref{sec:wscp} presents the weighted extension for covariate-shift settings. Sections~\ref{sec:sim} and~\ref{sec:app} evaluate the proposed method through simulations and a real data application, respectively. Section~\ref{sec:con} concludes with discussion and future directions.

\section{Structural Theory for SCP}\label{sec:scp}

\subsection{SCP Framework}
Let $\mathbf{X}$, $T$, $C$ denote the $p$-dimensional covariates, event time of interest and censoring time, respectively. Under general right censoring, we observe time $Y = \min(T, C)$ and the indicator $\delta = \mathbb{I}(T \leq C)$. Therefore, the data consist of $\{(\mathbf{X}_i, Y_i, \delta_i)\}_{i=1}^n$ where $n$ is the sample size. 
We assume conditionally independent censoring given the covariates, $T \indep C \mid \mathbf{X}$. 

We aim to construct a valid two-sided prediction interval $\widehat{C}(\mathbf{X}_{n+1})$ for the event time $T_{n+1}$ of a new subject with covariate $\mathbf{X}_{n+1}$, such that $\mathbb{P}\{T_{n+1} \in \widehat{C}(\mathbf{X}_{n+1}) ; \alpha\} \geq 1- \alpha$, for a target miscoverage rate $\alpha \in (0,1)$. 
Following the hypothesis testing interpretation of conformal prediction, each candidate time $t \in \mathcal{T}$ can be viewed as testing a null hypothesis $H_0: T_{n+1}=t$ \citep{shafer2008}. 
Let $\mathcal{D}_{tr}$ and $\mathcal{D}_{ca}$ denote the training and calibration sets, respectively. A nonconformity score $R(\mathbf{X}, t ; \mathcal{D}_{tr})$ is defined to measure how atypical a candidate time $t$ is for a subject with covariates $\mathbf{X}$, relative to the training data. In SCP, the quantile residual deviation is used as the nonconformity score \citep{romano2019}, 
\begin{align}
R(\mathbf{X},t;\mathcal{D}_{tr}) = \max\{\widehat{Q}_{\alpha/2}(\mathbf{X}; \mathcal{D}_{tr})-t, t-\widehat{Q}_{1-\alpha/2}(\mathbf{X}; \mathcal{D}_{tr})\},
    \label{eq:qresidual}
\end{align}
where $\widehat{Q}_{\alpha}(\mathbf{X}; \mathcal{D}_{tr})$ estimates the $\alpha$-th conditional quantile of $T \mid \mathbf{X}=\mathbf{x}$ .
Let $R_i=\delta_i R(\mathbf{X}_i, T_i ; \mathcal{D}_{{tr}}),\ i \in \mathcal{D}_{{ca}}$ denote the nonconformity scores for the calibration set, and $R_{n+1}(t)=R(\mathbf{X}_{n+1}, t ; \mathcal{D}_{\mathrm{tr}})$ be the score for a new subject at candidate time $t$. Because $R_i$ is only observed for uncensored cases, direct comparison with $R_{n+1}(t)$ is not possible. To address this, \cite{yietal2025} adopts a weight $w(\mathbf{X}, t) = f(\mathbf{X}, t)/f(\mathbf{X},t \mid\delta=1) \propto \{1-G(t\mid \mathbf{X})\}^{-1}$, where $f(\mathbf{X}, t)$ denotes the joint density of $(\mathbf{X},t)$, $f(\mathbf{X},t \mid\delta=1)$ the joint density for uncensored cases, and $G(t \mid \mathbf{X}) = \mathbb{P}(C \le t \mid \mathbf{X})$ the conditional censoring distribution. The censoring adjusted p-value is computed by 
\begin{align}
    p(t) = \sum_{i \in \mathcal{D}_{\mathrm{ca}} \cup \{n+1\}}
    \widetilde{w}_i(t)\,\mathbb{I}\{ R_i \ge R_{n+1}(t) \},
    \label{eq:pvalue}
\end{align}
where $\widetilde{w}_i(t)$ is the normalized weight of $w(\mathbf{X}_i, t)$,
\begin{align}
    \widetilde{w}_i(t) &= \frac{\delta_iw(\mathbf{X}_i,T_i)}{\sum_{j \in \mathcal{D}_{ca}}\delta_j w(\mathbf{X}_j, T_j) + w(\mathbf{X}_{n+1},t)} , i \in \mathcal{D}_{ca} \nonumber\\      \widetilde{w}_{n+1}(t) &= \frac{w(\mathbf{X}_{n+1},t)}{\sum_{j \in \mathcal{D}_{ca}}\delta_j w(\mathbf{X}_j, T_j) + w(\mathbf{X}_{n+1},t)}. \label{eq:SCPweight}
\end{align}
Then, the resulting prediction set is given by 
    $\widehat{C}(\mathbf{X}_{n+1}; \alpha) = \{ t \in \mathcal{T} : p(t) > \alpha \}.$

While this construction ensures marginal coverage, the structural properties of the resulting prediction interval have not been formally examined. In practice, SCP requires heuristic tuning choices--such as selecting censoring cutoffs and candidate time grids--and the resulting prediction sets, especially their upper bounds, can be highly sensitive to these choices under heavy censoring. These issues motivate the need for a more rigorous understanding of SCP and clearer guidance for constructing prediction set bounds.


\subsection{Quasi-Concavity of SCP p-values}\label{sec:pvalue}

To fill this gap, we first study the structural behavior of the SCP p-values introduced in Equation~\eqref{eq:pvalue}. In particular, we show that $p(t)$ is quasi-concave over $t\in \mathcal{T}$. The definition of quasi-concavity is as follows \citep{arrow1961, boyd2004, guerraggio2004}.




\begin{definition}
\label{def:quasi-concave}
A function $f: \mathcal{T} \to \mathbb{R}$ is quasi-concave if its upper level sets
    $A_c = \{t \in \mathcal{T}: f(t) \ge c\}$
are convex subsets of $\mathcal{T}$ for any $c \in \mathbb{R}$.
\end{definition}

This implies that if the upper level sets $\widehat{C}(\mathbf{X}_{n+1};\alpha)=\{t:p(t)>\alpha\}$ are convex subsets, they would form a single connected interval. In Proposition~\ref{prop:quasi-concavity}, we establish that the interval structure of the SCP prediction set is guaranteed by the quasi-concavity of $p(t)$. This result provides theoretical justification for defining the upper bound of the prediction interval as the first time $t$ at which $p(t)\leq \alpha$. 
 
\begin{proposition}[Quasi-concavity of SCP p-values]\label{prop:quasi-concavity}
The p-values $p(t)$ in Equation~\eqref{eq:pvalue} is quasi-concave in $t$. Therefore, the resulting conformal prediction set $\widehat C(\mathbf{X}_{n+1};\alpha)=\{t:p(t)>\alpha\}$ for any $\alpha\in(0,1)$ is an interval and connected.
\end{proposition}
Below, we present a brief sketch of proof that highlights the main ideas. A detailed proof of Proposition~\ref{prop:quasi-concavity} is available in Appendix~S1.1. 

\begin{proof}[Sketch of Proof for Proposition~\ref{prop:quasi-concavity}]
We establish the quasi-concavity of the p-value $p(t)$ in Equation~\eqref{eq:pvalue} by splitting the structural behavior into two components as
\begin{align}
     p(t) = \sum_{i \in \mathcal{D}_{\mathrm{ca}} \cup \{n+1\}}
    \widetilde{w}_i(t)\,\mathbb{I}\{ R_i \ge R_{n+1}(t) \} 
    = \sum_{i\in D_{ca}} \widetilde{w}_i(t) \mathbb{I}\{R_i\ge R_{n+1}(t)\} + \widetilde{w}_{n+1}(t).
\end{align}
We denote the normalized calibration contribution $\widetilde{p}_{\mathrm{ca}}(t) = \sum_{i\in D_{ca}} \widetilde{w}_i(t) \mathbb{I}\{R_i\ge R_{n+1}(t)\}$ and the test point weight term $\widetilde{w}_{n+1}(t)$.
\begin{enumerate}
    \item Calibration part: 
    We first define the unnormalized calibration p-value
    \begin{align}
        p_{\mathrm{ca}}(t)=\sum_{i\in\mathcal{D}_{ca}} w_i\, \mathbb{I}\{R_i \ge R_{n+1}(t)\}, 
        \label{eq:cali-pvalue}
    \end{align}
    where $w_i = \delta_i w(\mathbf{X}_i, T_i)$ does not depend on $t$.
    This separates the score comparison structure from the $t$-dependent normalization and test point weight, allowing us to analyze the intrinsic shape of the calibration contribution. Equation~\eqref{eq:qresidual} confirms that $\mathbb{I}\{R_i \ge R_{n+1}(t)\}$ depends only on the distance from the point
        $m := [\widehat{Q}_{\alpha/2}(\mathbf{X}_{n+1}; \mathcal{D}_{tr}) + \widehat{Q}_{1-\alpha/2}(\mathbf{X}_{n+1}; \mathcal{D}_{tr})]/2$. Thus, $p_{ca}(t)$ is non-increasing in the absolute deviation $|t - m|$. Then, we can define $\bar{s} = \inf \{|t-m| \ge 0: p_{ca}(|t-m|) < c\}$. This implies that for any $c\in \mathbb{R}$ its upper level set $\{t : p_{ca}(t) \ge c\} =\{t: |t-m| \le \bar{s} \}$ take the form of symmetric interval around $m$, showing that $p_{ca}(t)$ is a unimodal, quasi-concave function.\\
    
    \item Full weighted p-value: 
    The full p-value $p(t)$ is obtained by normalizing calibration p-value in Equation~\eqref{eq:cali-pvalue} and adding the test point weight term as $p(t)=\widetilde{p}_{\mathrm{ca}}(t)+\widetilde{w}_{n+1}(t)$.
    Since the weight $w(\mathbf{\mathbf{X}},t) \propto \{1-G(t\mid \mathbf{X})\}^{-1}$ is non-decreasing in $t$, both the normalization factor $\{\sum_{j \in D_{ca}}\delta_j w(\mathbf{X}_j, T_j) + w(\mathbf{X}_{n+1},t)\}^{-1}$ of $\widetilde{p}_{ca}(t)$ and the test point weight $\widetilde{w}_{n+1}(t)$ are monotone in $t$. Adding a monotone function to a unimodal, quasi-concave function preserves the convexity of all upper-level sets. Therefore, the full p-value $p(t)$ is also quasi-concave in $t$.
\end{enumerate}
Once quasi-concavity is established, it follows that the p-value can cross the significance level $\alpha$ at most twice. Consequently, the prediction set $\widehat{C}(\mathbf{X}_{n+1};\alpha)=\{t:p(t)>\alpha\}$ must be a single connected interval in $\mathcal{T}$.
\end{proof}
Beyond its theoretical significance, quasi-concavity plays an important diagnostic role. 
Proposition~\ref{prop:quasi-concavity} shows that the SCP p-value is structurally restricted to be a unimodal, quasi-concave function of the survival time $t$. 
Thus after $p(t)$ falls below a level $c\in\mathbb{R}$, it cannot rise above c again at any larger (or smaller) value of $t$ and 
must decay monotonically away from its peak. Therefore, any rebound or secondary increase in the tail constitutes a violation of quasi-concavity. Such tail instability tends to arise under heavy censoring due to identifiability issue, producing erratic prediction sets. These structural implications turn quasi-concavity into a practical diagnostic tool. When the p-value exhibits the expected quasi-concave shape, SCP yields a reliable two-sided interval whereas when quasi-concavity breaks down, the instability in the upper tail signals the need to switch to a more robust one-sided lower bound. 

\section{Weighted SCP under Covariate Shift}\label{sec:wscp}

\subsection{Weighted SCP}

The standard SCP framework is limited to settings where the training and test covariate distributions are identical. However, test populations often differ from the training population, which is a more realistic scenario known as covariate shift. We propose a weighted SCP method that extends SCP to account for potential covariate shift.

Suppose the original data consist of i.i.d. samples $\{(\mathbf{X}_i, T_i, \delta_i)\}_{i=1}^n$ drawn from $f^{original}=f_\mathbf{X} \times f_{T,C \mid \mathbf{X}}$. Let the new test data be $(\mathbf{X}_{n+1}, T_{n+1})$, where $T_{n+1}$ is unobserved. We assume a covariate shift setting where the test distribution factorizes as $f^{test} = f_\mathbf{X}^{test} \times f_{T,C \mid \mathbf{X}}$ with $f_\mathbf{X}^{test} \neq f_\mathbf{X}$ while the conditional event and censoring mechanisms $f_{T,C \mid \mathbf{X}}$ remain unchanged \citep{sugiyama2007, tibshiranietal2019}. Under this setting, our objective is to build data-driven prediction sets $\widehat{C}(\mathbf{X}_{n+1}; \alpha)$ that satisfy $\mathbb{P}\{T_{n+1} \in \widehat{C}(\mathbf{X}_{n+1})\} \geq 1- \alpha$ for $(\mathbf{X}_{n+1},T_{n+1})$ drawn from $f^{test}$.

We describe the procedures for the proposed weighted SCP framework as follows.

\vspace{0.3cm}
\noindent\textbf{Step 1: Define a nonconformity measure}\\
We divide the observed data into a training set $\mathcal{D}_{tr}$ for model fitting and a calibration set $\mathcal{D}_{ca}$ for conformity assessment. 
Using $\mathcal{D}_{tr}$, we estimate conditional quantiles of $T \mid \mathbf{X}$ and define the nonconformity score as in Equation~\eqref{eq:qresidual}.
In principle, our framework allows for any choice of nonconformity score. We adopt the quantile residual form in Equation~\eqref{eq:qresidual} because it performs favorably under heteroscedastic and skewed outcome distributions, which naturally arise in covariate dependent survival settings and typically yields shorter and more adaptive prediction intervals.

\vspace{0.3cm}
\noindent\textbf{Step 2: Adjust for censoring and covariate shift} \\
For each calibration subject, 
we compute $R_i = \delta_iR(\mathbf{X}_i, T_i;\mathcal{D}_{tr})$, and for the test case, $R_{n+1}(t)=R(\mathbf{X}_{n+1}, t; \mathcal{D}_{tr})$. 
Analogous to standard SCP, the calibration and test nonconformity scores are not directly comparable due to right censoring as well as the distributional discrepancy induced by covariate shift. To restore comparability, we apply a weighting strategy of \cite{tibshiranietal2019}, using a product of two density ratio terms. By the Bayes' rule, the weight can be written as
\begin{align}
    w^*(\mathbf{X},t) = \frac{f(\mathbf{X},t)}{f(\mathbf{X},t|\delta=1)} \frac{f^{test}(\mathbf{X},t)}{f(\mathbf{X},t)}\propto \frac{1}{1-G(t \mid \mathbf{X})}\frac{f_\mathbf{X}^{test}(\mathbf{X})}{f_\mathbf{X}(\mathbf{X})},
    \label{eq:newweight}
\end{align}
where $f_\mathbf{X}(\mathbf{X})$ and $f_\mathbf{X}^{test}(\mathbf{X})$ denote the marginal covariate densities of the original and test data, respectively. The first factor corrects for censoring via inverse probability weighting, and the second adjusts for the covariate distribution discrepancy between training and test samples. We define the weighted conformal p-value as
\begin{align}
    {p}^*(t) = \sum_{i \in \mathcal{D}_{ca} \cup \{n+1\}} \widetilde{w}_i^*(t)\mathbb{I}\{R_i \geq R_{n+1}(t)\},
    \label{eq:cov-pvalue}
\end{align}
where $\widetilde{w}_i^*(t)$ is obtained from the standard SCP weights in Equation~\eqref{eq:SCPweight} by replacing $w(\mathbf{X},t)$ with $w^*(\mathbf{X},t)$ in Equation~\eqref{eq:newweight}. 

\vspace{0.3cm}
\noindent\textbf{Step 3: Form the prediction set} \\
The final prediction interval becomes the collection of candidate times whose weighted p-values exceed the significance level $\alpha$:
    $\widehat{C}(\mathbf{X}_{n+1};\alpha) = \{t \in \mathcal{T}: {p}^*(t) > \alpha \}$.
A greater value of $p^*(t)$ indicates that the candidate time remains compatible with the weighted calibration distribution and therefore belongs to the prediction interval.

\subsection{Weighted Extension and Theoretical Preservation}

As defined in Equation~\eqref{eq:cov-pvalue}, the weighted p-value incorporates a density-ratio adjustment to correct for covariate shift. This raises a natural concern that the additional weighting could distort the shape of the p-value function and compromise the structural properties established in Section~\ref{sec:pvalue}. However, the covariate shift correction depends only on $\mathbf{X}$, not on the candidate time $t$, and therefore does not affect the time-varying structure of the p-value. The following Proposition~\ref{prop:preservation of qc} formalizes this.


\begin{proposition}[Preservation of Quasi-Concavity]\label{prop:preservation of qc}
For any $i$, the new weight in weighted SCP can be written as $w_i^*(\mathbf{X}_i,t)= w_i(\mathbf{X},t)\omega(\mathbf{X}_i)$, where $\omega(\mathbf{X}_i) \ge 0$, which is the density ratio and is independent of $t$, and $w_i(\mathbf{X},t)\ge 0$. Therefore, the quasi-concavity result of Proposition~\ref{prop:quasi-concavity} remains valid for the new weighted p-value ${p}^*(t)$ in Equation~\eqref{eq:cov-pvalue}.
\end{proposition}

The proof of Proposition~\ref{prop:preservation of qc} is available in Appendix~S1.2. This establishes that quasi-concavity--the structural property ensuring that the prediction set forms a single connected interval--remains intact after incorporating covariate shift adjustment.
Therefore, the interval shaped nature of SCP is preserved under covariate shift.

\subsection{Practical Implementation}\label{sec:implementation}

Having established that the weighted SCP preserves quasi-concavity, we discuss practical considerations for implementing the method. 
In the procedure of implementing the weighted SCP under covariate shift, three components need to be estimated: the conditional censoring distribution, the conditional quantiles for nonconformity score construction, and the covariate density ratio $f_X^{test}/f_X$. We describe each below.

\vspace{0.3cm}
\noindent\textbf{\boldmath Estimation of the $G(t\mid \mathbf{X})$} \quad
The inverse probability censoring weights involve $\{1- G(t\mid \mathbf{X})\}^{-1}$. 
We estimate $G(t \mid \mathbf{X})$ using a localized Kaplan-Meier estimator with Nadaraya-Watson kernel weights
\begin{align*}
    B_{nk}(\mathbf{x}) = \frac{K\left(\frac{\mathbf{x}-\mathbf{x}_k}{h_n}\right)}{\sum_{i=1}^n K\left(\frac{\mathbf{x}-\mathbf{x}_i}{h_n}\right)}
\end{align*}
where $K(\cdot)$ is a kernel density function with bandwidth $h_n \in \mathbb{R}^+$ converging to zero as $n \to \infty$. Using these weights, the conditional censoring distribution is estimated by 
\begin{align*}
    \widehat{G}(t \mid \mathbf{X}) = 1- \prod_{j=1}^n \left\{1- \frac{B_{nj}(\mathbf{X})}{\sum_{k=1}^n \mathbb{I}(Y_k \ge Y_j)B_{nk}(\mathbf{X})}\right\}^{\eta_j(t)}
\end{align*}
where $\eta_j(t) = \mathbb{I}(Y_j \le t, \delta_j =0)$. 
In practice, for numerical stability, we truncate $\widehat{G}(t\mid\mathbf{X})$ so that $\widehat{G}(t\mid\mathbf{X})<1$, ensuring that the weight $\{1-\widehat{G}(t\mid\mathbf{X})\}^{-1}$ remains finite. 

\vspace{0.3cm}
\noindent\textbf{Estimation of conditional quantiles} \quad
Accurate and identifiable conditional quantiles, especially the upper quantile $Q_{1-\alpha/2}(\mathbf{X})$, are important for maintaining stability in the right tail of $p^*(t)$.
To estimate conditional quantiles under right censoring, we adopt the locally weighted censored quantile regression method, which incorporates censored observations through redistribution-of-mass \citep{wang2009}. 
We note that upper conditional quantiles can become non-identifiable under heavy censoring where the largest observable event times are often censored. 
Identifiability requires $G(Q_{1-\alpha/2}(\mathbf{X})\mid\mathbf{X})<1$, on a set of covariate values with positive measure. When this condition fails, the upper quantile estimate becomes unstable, and this instability directly propagates to the right tail of ${p}^*(t)$, producing violations of the quasi-concavity structure that determine the interval's reliability. 

\vspace{0.3cm}
\noindent\textbf{\boldmath Estimation of density ratio} \quad
To adjust for covariate shift, we estimate the likelihood ratio $\omega(\mathbf{X}) =f_X^{test}/f_X$ using a probabilistic classifier of test and training covariates. In the proposed weighted SCP, the estimated probability $\hat{p}(\mathbf{X})$ is obtained by a random forest classifier trained to predict whether a covariate belongs to the test distribution. Following \cite{sugiyama2012} and \cite{tibshiranietal2019}, the density ratio can be expressed through the conditional odds
\begin{align*}
    \hat{\omega}(\mathbf{X}) = \frac{\hat{p}(\mathbf{X})}{1-\hat{p}(\mathbf{X})}.
\end{align*}
Because excessively large density ratios may cause a few observations to dominate the weighted $p$-value $p^*(t)$ and thereby introduce numerical instability, we apply clipping to $\hat{\omega}(\mathbf{X})$, by truncating values that fall outside a 0.01 and 0.99 range. 
By truncating, we prevent $\hat{\omega}(\mathbf{X})$ from being infinite. 


\section{Simulation}\label{sec:sim}

We conduct a series of simulation studies to evaluate whether the proposed weighted SCP method achieves the target coverage $1-\alpha$ across various settings. We examine the performance under four scenarios formed by combining two data-generating mechanism (i.e., linear homoscedastic vs. linear heteroscedastic) with two covariate-distribution settings (i.e., with vs. without covariate shift). We consider sample sizes $n=300$ and $800$, censoring rates of $20, 40, 60$, and $80\%$. Each experiment is replicated for $500$ times. In each replication, we independently generate a test set of size $100$.

In simulation, we compare the proposed weighted SCP with the standard SCP framework of \cite{yietal2025} using average coverage (AC) and average interval length (AL) of the prediction sets as evaluation metrics. We set $\alpha=0.1$ as the miscoverage level, corresponding to a target coverage of $90\%$.

\subsection{Simulation Settings}

We generate two-dimensional covariates $\mathbf{X}_i = (X_{i1}, X_{i2})^\top$ with each element independently drawn from a uniform distribution $U(0,1)$. We assume that survival times follow the model, $\log(T_i) = \beta_0 + \mathbf{X}_i^T \boldsymbol{\beta} + \epsilon_i,$ where $\beta_0 =2$ and coefficients $\boldsymbol{\beta} = (3, -1)^T$, for $i=1, \dots ,n$. We consider two error distributions as follows.
\begin{enumerate}
    \item Linear homoscedastic model: $\epsilon_i \sim N(0,0.5)$
    \item Linear heteroscedastic model: $\epsilon_i \sim \text{Gamma}(0.5|X_{i1}|, 0.3 + 5|X_{i2}|)$.
\end{enumerate}

We also consider two covariate-distribution settings to evaluate the validity and robustness of the proposed method as follows.
\begin{enumerate}
    \item With covariate shift: We impose covariate shift by exponential tilting where $25\%$ of test covariates are resampled with probability proportional to $\omega(\mathbf{X}_i) = \exp(\mathbf{X}_i^T \boldsymbol{\gamma})$, where $\boldsymbol{\gamma}=(-1, 1)^T$. We note that $\omega(\mathbf{X}_i)$ is a density ratio of covariate distributions between training and test sets. This construction only modifies the marginal covariate distribution while leaving the conditional survival and censoring mechanisms unchanged. 
    \item Without covariate shift: We generate the test set from the same distribution as the training set.
\end{enumerate}

Censoring times $C_i$ are further generated from $U(X_{i1} + X_{i2}, a_0 + X_{i1} + X_{i2})$,
where constant $a_0=(10, \ 4.75, \ 3, \ 2.04)$ for the homoscedastic model and $a_0=(8.3, \ 4.4, \ 3.15, \ 2.15)$ for the heteroscedastic model, yielding censoring rates of approximately $20, 40, 60$, and $80\%$, respectively. 




When implementing the proposed method, kernel smoothing is required for estimating both conditional censoring distribution and conditional quantile regression. To ensure comparable smoothing across different levels of censoring, we determine the bandwidths based on the effective sample size associated with each estimation task. For estimating $G(t \mid \mathbf{X})$, the effective sample size is defined as $n_{\mathrm{eff}} = n \times (\text{proportion of censored cases})$, while for estimating $F(t \mid \mathbf{X})$, we use $n_{\mathrm{eff}} = n \times (\text{proportion of uncensored cases})$. Following \cite{wang2009} and \cite{yietal2025}, we choose the bandwidth $h_n$ according to
\begin{align*}
    h_n =
\begin{cases}
0.50, & n_{\mathrm{eff}} \le 200, \\
0.35, & 200 < n_{\mathrm{eff}} \le 400, \\
0.25, & n_{\mathrm{eff}} > 400.
\end{cases}
\end{align*}

\subsection{Simulation Results}

Table~\ref{tab:scenario1} summarizes the simulation results under covariate shift. The proposed weighted SCP method consistently achieves coverage close to the nominal $90\%$ level across all censoring rates and both error structures. Specifically, in the homoscedastic setting, the coverage ranges from 90.19 to 94.39\% for $n=300$ and from 89.80 to 93.56\% for $n=800$, while interval lengths increase with the censoring rates. In contrast, the standard SCP exhibits noticeable undercoverage under heavy censoring rates of 60 or 80\%, especially for $n=800$, and its interval lengths do not always increase monotonically with censoring. Similar patterns are observed under heteroscedastic errors, where the proposed weighted SCP maintains valid coverage, albeit slightly over covered under a heavy censoring rate of 80\%; whereas standard SCP suffers undercoverage in the same settings with no consistent patterns in interval lengths.


\begin{table}[htbp]%
\caption{Average coverage (AC) and average interval length (AL) under the covariate shift setting.\label{tab:scenario1}}
{\tabcolsep0pt\begin{tabular*}{\textwidth}{@{\extracolsep{\fill}}l c c c c c c c c c@{}}
\toprule
 & & \multicolumn{4}{c}{\textbf{Homoscedastic}} & \multicolumn{4}{c}{\textbf{Heteroscedastic}} \\
\cmidrule(lr){3-6} \cmidrule(lr){7-10}
 & &\multicolumn{2}{@{}c@{}}{\textbf{Weighted SCP}} & \multicolumn{2}{@{}c@{}}{\textbf{SCP}}&\multicolumn{2}{@{}c@{}}{\textbf{Weighted SCP}} & \multicolumn{2}{@{}c@{}}{\textbf{SCP}} \\
\cmidrule{3-4}\cmidrule{5-6}\cmidrule{7-8}\cmidrule{9-10}
$\textbf{n}$ & \textbf{Cenoring rate} & \textbf{AC$\%$} & \textbf{AL} & \multicolumn{1}{@{}l@{}}{\textbf{AC$\%$}} & \textbf{AL} & \textbf{AC$\%$} & \textbf{AL} & \multicolumn{1}{@{}l@{}}{\textbf{AC$\%$}} & \textbf{AL} \\
\midrule
300 & 20$\%$ & 90.61 & 72.91 & 91.86 & 63.62 & 90.60 & 69.95 & 91.68 & 88.67\\
& 40$\%$ & 90.19 & 92.44 & 90.11 & 56.66 & 89.03 & 95.25 & 87.89 & 45.74\\
& 60$\%$ & 92.61 & 111.44 & 84.29 & 37.53 & 91.07 & 107.95 & 83.68 & 26.62\\
& 80$\%$ & 94.39 & 113.28 & 88.37 & 77.64 & 95.97 & 111.23 & 88.58 & 144.81 \\
\midrule
800 & 20$\%$ & 90.00 & 57.77 & 91.64 & 58.83 & 89.94 & 47.32 & 89.23 & 44.81\\
& 40$\%$ & 89.80 & 66.07 & 90.44 & 51.86 & 89.14 & 59.47 & 86.81 & 31.68 \\
& 60$\%$ & 91.47 & 96.74 & 81.60 & 31.14 & 92.26 & 90.32 & 80.85 & 19.49 \\
& 80$\%$ & 93.56 & 107.09 & 69.65 & 19.65 & 95.18 & 104.02 & 73.29 & 13.67\\
\bottomrule
\end{tabular*}} 
\end{table}

Table~\ref{tab:scenario2} reports results when no covariate shift is present. Even when the weighted method is applied unncessarily, assuming there exist discrepancy in covariate distributions between training and test sets, it remain robust. Overall coverage of the proposed method is close to the nominal level and comparable to the standard SCP approach across all settings for $n=300$. This demonstrates that the proposed weighted SCP has no loss of performance when covariate distributions are identical. 

\begin{table}[htbp]%
\caption{Average coverage (AC) and average interval length (AL) under no covariate shift setting.\label{tab:scenario2}}
{\tabcolsep0pt\begin{tabular*}{\textwidth}{@{\extracolsep{\fill}}l c c c c c c c c c@{}}
\toprule
 & & \multicolumn{4}{c}{\textbf{Homoscedastic}} & \multicolumn{4}{c}{\textbf{Heteroscedastic}} \\
\cmidrule(lr){3-6} \cmidrule(lr){7-10}
 & &\multicolumn{2}{@{}c@{}}{\textbf{Weighted SCP}} & \multicolumn{2}{@{}c@{}}{\textbf{SCP}}&\multicolumn{2}{@{}c@{}}{\textbf{Weighted SCP}} & \multicolumn{2}{@{}c@{}}{\textbf{SCP}} \\
\cmidrule{3-4}\cmidrule{5-6}\cmidrule{7-8}\cmidrule{9-10}
$\textbf{n}$ & \textbf{Cenoring rate} & \textbf{AC$\%$} & \textbf{AL} & \multicolumn{1}{@{}l@{}}{\textbf{AC$\%$}} & \textbf{AL} & \textbf{AC$\%$} & \textbf{AL} & \multicolumn{1}{@{}l@{}}{\textbf{AC$\%$}} & \textbf{AL} \\
\midrule
300 & 20$\%$ & 89.23 & 77.39 & 91.77 & 77.93 & 89.65 & 72.31 & 91.81 & 66.14\\
& 40$\%$ & 89.75 & 94.78 & 91.11 & 80.60 & 89.97 & 95.56 & 90.67 & 70.62\\
& 60$\%$ & 91.84 & 110.41 & 89.90 & 98.16 & 91.81 & 105.98 & 89.26 & 80.20\\
& 80$\%$ & 93.50 & 112.92 & 93.24 & 112.29 & 95.39 & 110.82 & 94.63 & 108.27\\
\bottomrule
\end{tabular*}} 
\end{table}

\section{Real Data Application}\label{sec:app}

We applied the proposed framework to two publicly available cohorts of node-positive breast cancer patients: the Rotterdam Tumor Bank and the German Breast Cancer Study Group (GBSG). Both datasets record recurrence-free survival following primary surgery and are widely used as benchmarks for evaluating survival prediction methods. In our application, the Rotterdam data were used as the training and calibration sets, and the GBSG data served as an external test set.

Following the established criteria ensuring a comparable patient sample across the two cohorts \citep{roystonetal2013,craigetal2025}, we restricted the analysis to subjects with at least one positive lymph node, yielding 1,546 patients from the Rotterdam cohort and 686 from the GBSG cohort. The two datasets share seven common covariates: patient's age at surgery, tumor grade, number of positive lymph nodes, progesterone receptor level, estrogen receptor level, and binary indicators of hormonal treatment and menopause status. Recurrence-free survival time measured in months, after converting the original day based measurements, was used as the outcome, with the event indicator taking the value of 1 for recurrence or death from any cause and 0 for censoring. The censoring rates were approximately $37\%$ and $56\%$, for the Rotterdam and GBSG data, respectively.

To investigate whether a covariate shift is present between the training and test data, we compared the marginal distributions of the shared covariates. As shown in Table~\ref{tab:baseline_table}, notable differences exist for several key covariates, including age, tumor grade, hormonal treatment, and menopause status. This presents evidence that the real-world setting involves a meaningful shift in covariate distribution, motivating the use of the proposed weighted SCP method to adjust for distributional differences between training and test data.

\begin{table}[htbp]
\caption{Baseline characteristics of node-positive breast cancer patients in the Rotterdam (training) and GBSG (test) datasets.\label{tab:baseline_table}}
{\tabcolsep0pt
\begin{tabular*}{\textwidth}{@{\extracolsep{\fill}}lcc@{}}
\toprule
\textbf{Variable} & \textbf{Rotterdam} & \textbf{GBSG} \\
\midrule
n & 1546 & 686 \\
Censoring rate & 37\% & 56\% \\
\midrule
\multicolumn{3}{@{}l}{\textbf{Continuous covariates, Mean (SD)}}\\[3pt]
Age   & 55.98 (13.00)   & 53.05 (10.12) \\
Estrogen receptor level    & 165.08 (267.28) & 96.25 (153.08) \\
Progesterone receptor level   & 156.22 (299.44) & 110.00 (202.33) \\
Positive lymph nodes & 5.23 (4.89)     & 5.01 (5.48) \\
\midrule
\multicolumn{3}{@{}l}{\textbf{Categorical covariates, n (\%)}}\\[3pt]

Tumor grade & & \\
\hspace{1em}Grade 1 & 0 (0.0\%)     & 81 (11.8\%)  \\
\hspace{1em}Grade 2 & 369 (23.9\%)  & 444 (64.7\%) \\
\hspace{1em}Grade 3 & 1177 (76.1\%) & 161 (23.5\%) \\[3pt]

Hormonal treatment & & \\
\hspace{1em}Not received & 1207 (78.1\%) & 440 (64.1\%) \\
\hspace{1em}Received     & 339 (21.9\%)  & 246 (35.9\%) \\[3pt]

Menopause status & & \\
\hspace{1em}Pre-menopause  & 628 (40.6\%) & 290 (42.3\%) \\
\hspace{1em}Post-menopause & 918 (59.4\%) & 396 (57.7\%) \\
\bottomrule
\end{tabular*}}
\end{table}

The proposed weighted SCP method was applied to construct the prediction sets. Direct evaluation of empirical coverage is not possible for real censored data, as their true event times are unobserved. Thus, average coverage is computed using only uncensored test subjects, whereas the average interval length is calculated for all observations. Table~\ref{tab:realdata_wscp} summarizes the average coverage and average interval length for both the proposed weighted SCP and the standard SCP on the GBSG test set. The weighted SCP method achieves an empirical coverage of $93.31\%$, which is close to the nominal level. This coverage rate is substantially higher than the standard SCP ($83.61\%$), indicating that adjusting for covariate shift leads to more reliable prediction intervals. Notably, this improvement is obtained with only a modest increase in interval width, suggesting that the density ratio weighting approach enhances coverage without excessively inflating the length of the prediction intervals.

\begin{table}[htbp]
\caption{Average coverage (AC) and average interval length (AL) for the GBSG test data.\label{tab:realdata_wscp}}
{\tabcolsep0pt
\begin{tabular*}{\textwidth}{@{\extracolsep{\fill}}lcc@{}}
\toprule
\textbf{Method} & \textbf{AC\%}$^\dagger$ & \textbf{AL} \\
\midrule
Weighted SCP
    & 93.31\% & 164.545\\[3pt]
SCP
    & 83.61\% & 141.933\\ 
\bottomrule
\end{tabular*}}
\\[3pt]
{\footnotesize
$^\dagger$Average coverage is computed only using uncensored test points} 
\end{table}


\section{Conclusion}\label{sec:con}

This work develops a theoretical and methodological foundation for survival conformal prediction under right censoring and covariate shift. We first establish that the SCP conformal p-value is quasi-concave in the candidate time, guaranteeing that the resulting prediction set is always a single, connected interval. This result formalizes a structural property that had previously been supported only empirically and provides new insight into how censoring affects the shape of the conformal p-value, including when the upper tail instability may arise due to identifiability limitations. Building on this foundation, we extend SCP to setting with covariate shift using a density ratio weighting scheme. Because the additional weighting depends only on the covariates and not on the candidate time, we show that the key structural properties of SCP are preserved under shift.

Our empirical studies further demonstrate the practical value of the proposed framework. Simulations confirm that covariate shift weighting substantially improves empirical coverage when the test covariate distribution differs from the calibration distribution, while maintaining stable interval lengths. An application to the Rotterdam and GBSG breast cancer datasets illustrates robustness in real survival data with heterogeneous covariates and nontrivial shift in their distributions. 

Future work include analyzing the sensitivity of coverage to density ratio estimation under extreme covariate shift settings, extending the method to sequential or online prediction settings, and conducting broader validation across diverse clinical datasets. These developments will further strengthen survival conformal prediction as a principled and interpretable uncertainty-quantification tool for time-to-event outcomes.


\section*{Acknowledgments}
This work was supported by the National Research Foundation of Korea (NRF) grant funded by the Korea government (MSIT) [RS-2024-00341883], and by the Yonsei University Research Fund of 2025 [2025-22-0429].

\beginsupplement
\section{Appendix}
\subsection{Proof of Proposition~1}
\begin{proof}[Proof of Proposition~1] \quad
	Let $a = \widehat{Q}_{\alpha/2}(\mathbf{X}_{n+1}; \mathcal{D}_{tr})$, $b =\widehat{Q}_{1-\alpha/2}(\mathbf{X}_{n+1}; \mathcal{D}_{tr})$, $a\leq b$ denote the estimated lower and upper quantiles from the training set $D_{tr}$ and define their midpoint by $m = (a+b)/2$. Then, the test nonconformity score is 
	\begin{align*}
		R_{n+1}(t) &= R(\mathbf{X}_{n+1}, t; D_{tr})\\ 
		&= \text{max}\{\widehat{Q}_{\alpha/2}(\mathbf{X}_{n+1};D_{tr})-t, t-\widehat{Q}_{1-\alpha/2}(\mathbf{X}_{n+1};D_{tr})\}\\
		&= \text{max}\{a-t, t-b\}.
	\end{align*}
	Since the $R_{n+1}(t)$ is the maximum of the two affine functions, it is convex and attains its unique minimum at $t=m$.
	Let $r_i = R_i \geq 0$ denote the calibration nonconformity score, for $i \in D_{ca}$. The set $\{R_i \geq R_{n+1}(t) \}$ is $\{R_i \geq \max(a-t, t-b)\} = \{\max(a-t, t-b) \leq r_i\} = \{a-r_i \leq t \leq b+r_i\}$.
	Since $m = (a+b)/2$, the above set can be written as $\{a-r_i \leq t \leq b+r_i\} = \{|t-m| \leq s_i\}$ where $s_i=(b-a)/2 + r_i \geq 0$.
	Thus each indicator has the form
	\begin{align}
		\mathbb{I}\{R_i \geq R_{n+1}(t)\} = \mathbb{I}\{|t-m| \leq s_i\}. \tag{1} \label{eq:indicator}
	\end{align}
	First define the unnormalized weighted p-value using the calibration set
	\begin{align*}
		p_{ca}(t) = \sum_{i \in D_{ca}} w_i \mathbb{I}\{R_i \geq R_{n+1}(t)\}, \ \ w_i \geq 0.
	\end{align*}
	Using \eqref{eq:indicator}, we obtain
	\begin{align*}
		p_{ca}(t) = \sum_{i \in D_{ca}} w_i \mathbb{I}\{|t-m| \leq s_i\}.
	\end{align*}
	Let $d = |t-m|$. Then
	\begin{align*}
		p_{ca}(t) = H(d) = \sum_{i \in D_{ca}} w_i \mathbb{I}\{d \leq s_i\}.
	\end{align*}
	As $d$ increases, the set $\{i : d \leq s_i\}$ shrinks. Thus $H(d)$ is non-increasing in $d$. Hence for any $c>0$, define the threshold $\bar{s} = \inf \{d \geq 0 : H(d) < c\}$.
	The upper level set of $p_{ca}(t)$ is
	\begin{align*}
		\{t : p_{ca}(t) \geq c\} = \{t: |t-m| \leq \bar{s}\} = [m-\bar{s}, m+\bar{s}].
	\end{align*}
	Because the upper level set is always an interval, $p_{ca}(t)$ is quasi-concave and unimodal, achieving its maximum at $t=m$. \citep{arrow1961, boyd2004, guerraggio2004}\\
	The calibration part of the SCP p-value is
	\begin{align*}
		\tilde{p}_{ca}(t) = \sum_{i \in D_{ca}}\widetilde{w}_i(t) \mathbb{I}\{R_i \geq R_{n+1}(t)\} = \zeta(t)p_{ca}(t),
	\end{align*}
	where
	\begin{align*}
		\zeta(t) = \frac{1}{\sum_{j \in D_{ca}}\delta_jw(\mathbf{X}_j, T_j)+w(\mathbf{X}_{n+1},t)}.
	\end{align*}
	Since $w(\mathbf{X}, t) \propto (1-G(t \mid \mathbf{X}))^{-1}$, $w(\mathbf{X},t)$ is non-decreasing in $t$, and $\zeta(t)$ is non-increasing in $t$. Thus $\zeta(t)$ acts as a positive monotone scaling factor. Multiplying a quasi-concave function by a positive monotone factor cannot create new local maxima or disconnect the upper level set. Indeed, for any $c^* >0$,
	\begin{align}
		\tilde{p}_{ca}(t) \ge c^* \quad \Leftrightarrow \quad {p}_{ca}(t) \ge \frac{c^*}{\zeta(t)}. \tag{2} \label{eq:norm-cali-pvalue}
	\end{align}
	Because $\zeta(t)$ is monotone and ${p}_{ca}(t)$ has convex upper level set, the right hand side of \eqref{eq:norm-cali-pvalue} also defines an interval in $t$. Therefore $\tilde{p}_{ca}(t)$ is quasi-concave. \\
	The full p-value in SCP framework is
	\begin{align*}
		p(t) = \tilde{p}_{ca}(t) + \widetilde{w}_{n+1}(t),
	\end{align*}
	where
	\begin{align*}
		\widetilde{w}_{n+1}(t) = \frac{w(\mathbf{X}_{n+1}, t)}{\sum_{j \in D_{ca}} \delta_j w(\mathbf{X}_j, T_j) + w(\mathbf{X}_{n+1}, t)}
	\end{align*}
	is monotone increasing in $t$. For any $c^{**} >0$,
	\begin{align*}
		\{t: p(t) \ge c^{**}\} = \{t: \tilde{p}_{ca}(t) \ge c^{**} - \widetilde{w}_{n+1}(t)\}. \tag{3} \label{eq:full-pvalue}
	\end{align*}
	Since $\tilde{p}_{ca}(t)$ is quasi-concave and unimodal in $t$, and $\widetilde{w}_{n+1}(t)$ is monotone, the solution of \eqref{eq:full-pvalue} cannot split into multiple disjoint intervals. Hence every upper level set $\{t: p(t) \ge c^{**}\}$ is an interval, thus $p(t)$ is quasi-concave.
\end{proof}

\subsection{Proof of Proposition~2}
\begin{proof}[Proof of Proposition~2] \quad
	By Proposition~1, the SCP p-value $p(t)$ is quasi-concave in $t$. Therefore, to establish quasi-concavity of the covariate shifted p-value ${p}^*(t)$, it is
	sufficient to verify that the additional weighting factor introduced by the covariate shift enters only through a nonnegative multiplier that is independent of $t$.
	The weighting function can be decomposed as
	\begin{align*}
		w^*(\mathbf{X},t)
		&= \frac{f^{test}(\mathbf{X},t)}{f(\mathbf{X},t)} \cdot \frac{f(\mathbf{X},t)}{f(\mathbf{X},t \mid \delta=1)} 
		= \frac{f_X^{test}(\mathbf{X})}{f_X(\mathbf{X})}
		\cdot \frac{f(\mathbf{X},t)}{f(\mathbf{X},t \mid \delta=1)} 
		\propto \frac{\hat{p}(\mathbf{X})}{1-\hat{p}(\mathbf{X})} \cdot \frac{1}{1-G(t\mid \mathbf{X})},
	\end{align*}
	where $\hat{p}(\mathbf{X})$ denotes an estimate of $\mathbb{P}(Z=1 \mid \mathbf{X}=\mathbf{x})$ obtained from a binary classifier distinguishing test points (Z=1) from training points (Z=0). The covariate shift factor
	\begin{align*}
		\omega(\mathbf{X})=\frac{f_X^{test}(\mathbf{X})}{f_X(\mathbf{X})}
	\end{align*}
	is nonnegative and independent of $t$, so the new weights take the form
	\begin{align*}
		w_i^*(\mathbf{X},t)= w_i(\mathbf{X},t)\omega(\mathbf{X}),
	\end{align*}
	which satisfies the stated conditions. Multiplication by a nonnegative constant independent of $t$ does not change the upper-level sets of $p(t)$, and therefore does not affect quasi-concavity. Hence, the weighted p-value ${p}^*(t)$ inherits the quasi-concavity of $p(t)$.
\end{proof}

\newpage
\bibliographystyle{apalike} 
\bibliography{ref}

\end{document}